# Direct evidence of the orbital contribution to the magnetic moment in AA'FeReO$_6$ double perovskites

M. Sikora[1,2], D. Zajac[1,2], Cz. Kapusta[1], M. Borowiec[1], C.J. Oates[1], V. Prochazka[1,3], D. Rybicki[1], J.M. De Teresa[2], C. Marquina[2] and M.R. Ibarra[2]

[1] *Department of Solid State Physics, Faculty of Physics and Applied Computer Sciences, AGH University of Science and Technology, 30-059 Cracow, Poland*

[2] *Departamento de Fisica de la Materia Condensada and Instituto de Ciencia de Materiales de Aragon, Universidad de Zaragoza - CSIC, 50009 Zaragoza, Spain*

[3] *Faculty of Mathematics and Physics, Charles University, V Holesovickach 2, 180 00 Prague 8, Czech Republic*

**Abstract**

Spin and orbital magnetic moments of Re in AA'FeReO$_6$ double perovskites (A,A' = Ba, Sr, Ca) have been directly probed employing XMCD spectroscopy at the Re L$_{2,3}$-edges. A considerable orbital magnetic moment is observed in all the studied compounds despite octahedral coordination. Relative orbital to spin contribution per Re atom rises with lattice distortion from $m_L/m_S$ = -0.28 to -0.34 for AA'=Ba$_2$FeReO$_6$ and Ca$_2$FeReO$_6$, respectively. A preliminary XMCD measurements at the Fe L$_{2,3}$-edges reveals also a significant orbital moment of iron in Ca$_2$FeReO$_6$. The relation of the results to the magnetic properties of the compounds is discussed.

Ordered double perovskites has recently attracted great interest due to their large spin polarization and Curie temperature (Tc) much higher than room temperature. These properties are strongly desired in order to realize reasonable magnetoresistance effects at room temperature, which is not only challenging subject of fundamental science but also important phenomenon for potential application in spin electronics. Therefore the first observation of substantial magnetoresistance at room temperature in Sr$_2$FeMoO$_6$ [1] quickly led to production of new devices like magnetic tunnel junctions and magnetoresistive potentiometers.

The ferromagnetic metallic properties of Sr$_2$FeMoO$_6$ have been explained by the indirect ferromagnetic interaction Fe-O-Mo. Band calculations and optical measurements indicate that the spin-up subband is below the Fermi level and consists of five "localized" Fe 3d electrons. The spin-down subband would be located at the Fermi level and would consist of one "delocalized" electron shared by Fe and Mo and mediating the ferromagnetic interaction between the "localized" spins through the oxygen orbitals [1]. Ideally, the conduction electrons would show complete negative spin polarization at the Fermi level, i.e. half-metallicity. In order to proof the relevance of the proposed model, the measurements of magnetic moment at the Mo site have been performed by means of neutron diffraction [2], X-ray Magnetic Circular Dichroism (XMCD) [3,4] and Nuclear Magnetic Resonance (NMR) [5] studies. Although the first XMCD measurements [3] did not reveal reasonable value of Mo magnetic moment, the contradictory results were obtained in another XMCD study [4] as well as in NMR and neutron diffraction experiments, which proved that Mo bears a magnetic moment of order of $0.5\mu_B$ antiparallel to the Fe moment, as the predicted theoretically.

Currently the other ordered double perovskites $A_2BB'O_6$ (A = Ca, Sr, Ba, La, etc.; BB' = FeMo, FeRe, CrRe, CrW, etc.) are being intensively studied in order to find a material with optimal performance [6-11]. Among them the Re-based double perovskites are the most promising compounds in terms of high Curie temperature, e.g. $T_C$ = 538K and 635K in $Ca_2FeReO_6$ and $Sr_2CrReO_6$, respectively [7]. However, these compounds show even a more variety of magnetic and transport properties than those of Mo-based double perovskites due to an additional electron in spin-down subband and stronger spin-orbit coupling in heavier Re ion [7-11]. Although the spin moment of the additional electron decreases bulk magnetic moment from $\sim 4\mu_B$ in $AA'FeMoO_6$ to $\sim 3\mu_B$ in $AA'FeReO_6$ double perovskites, its influence on the strength of ferromagnetic interaction seems to be weak, since Curie temperatures of $Sr_2FeMoO_6$ and $Sr_2FeReO_6$ are almost identical, 410K and 400K, respectively. Peculiar differences are observed in evolution of the transport properties caused by decreasing of average ionic radius ($<r_A>$) of alkali earth ions. Smaller A-site ions lead to lattice distortion and lowering of a local symmetry from cubic (AA'=$Ba_2$) to tetragonal (AA'=$Sr_2$) and monoclinic/orthorhombic (AA'=$Ca_2$) in Re/Mo-based compounds. Then, the Mo-based compounds preserves its weak metallicity, whereas Re-based double perovskites of monoclinic symmetry shows semiconducting behavior and undergoes a structural transition at low temperatures, which further increases its resistivity [7,9,10,11]. Especially intriguing is the difference in the magnetic behavior. The Re-based double perovskites are magnetically hard [8,10,12] and reveal large magneto-elastic effects [13], which can only be explained by large magnetorystalline anisotropy due to not fully quenched orbital moment of Re [10]. This assumption is also supported by the results of recent self-consistent LSDA and GGA calculations, which revealed significant value of $m_L$=$0.25\mu_B$ per Re atom in $Sr_2FeReO_6$ [14].

The main aim of present study is to find out, whether a large nonquenched orbital moment exist in Re-based double perovskites and how lattice distortion influences its value. Moreover the certain information on total magnetic moment of Re ions and its average valence states need to be probed more precisely, since those measured by neutron diffraction [9,15], Mösbauer [6,8,16] and NMR [17] spectroscopies as well as calculated by spin density method [14,18,19] are consistent only qualitatively. For example, the calculated and measured magnetic moment of Re is always antiparallel to that of Fe, but their values are estimated in the range from $0.53\mu_B$ to $1.33\mu_B$, whereas the valence state is assumed to be between $Re^{5+}$ ($5d^2$) and $Re^{6+}$ ($5d^1$).

We address all these issues utilizing Sum Rules analysis to the XMCD spectra at Re $L_{2,3}$-edges, which enables us to derive the spin and orbital contributions to the magnetic moment independently [20]. Although Sum

Rules may give unreliable results, when applied to the 3d transition metals, due to overlap of absorption edges, its application to analysis of 5d spin and orbital moments, induced by iron in Fe/5d multilayers, showed results in excellent agreement with relativistic spin-polarized LMTO calculations [21]. The applicability of XMCD to study 5d magnetic moments in ferromagnetic alloys was also proved in recent experiments on IrMnAl system [22].

The experiments were carried out on the series of five AA'FeReO$_6$ double perovskites (AA' = Ba$_2$, BaSr, Sr$_2$, SrCa, Ca$_2$). Samples were prepared by solid state reaction of the A$_2$CO$_3$, Fe$_2$O$_3$, ReO$_3$ and pure Re (the ReO$_3$/Re ratio was 5/1) at 1000 °C during 3 h in an atmosphere of Ar. Detailed description of the preparation route as well as structural, transport and magnetic properties of the compounds are described in ref. [18]. XMCD measurements at Re L$_{2,3}$-edges were performed at the A1 beamline in Hasylab/DESY, Hamburg. The spectra were recorded at 10K for polycrystalline samples using transmission mode in an applied magnetic fields of 5 or 20kOe, flipped at each energy point. The degree of circular polarization of the bending magnet radiation was estimated to $P_C$ = 77% and 79% at L$_3$ and L$_2$-edge, respectively, behind the double crystal Si(111) monochromator. All the spectra have been measured eight to ten times using two opposite sequence of magnetic field flipping (NSSN, SNNS) in order to minimize systematic errors. Absorption spectra were background subtracted and normalized to unity at L$_3$-edge, whereas XMCD spectra at were normalized to the edge step, P$_C$ and relative magnetization. XMCD spectra at Fe L$_{2,3}$-edges were performed at the BACH beamline of the Elettra Synchrotron Laboratory in Trieste. Circularly polarized radiation was produced by a helical undulator ($P_C$ > 99.5%) and monochromatized by spherical gratings. The spectra were recorded at room temperature for bulk polycrystalline samples of BaSrFeReO$_6$, Sr$_2$FeReO$_6$ and Ca$_2$FeReO$_6$ using total electron yield detection technique. Four consecutive XAS measurements, two for each helicity, were performed on samples magnetized ex situ to 50 kOe and placed on a small Nd-Fe-B permanent magnet.

Re L$_{2,3}$ absorption spectra of Ba$_2$FeReO$_6$ presented in fig.1 have a shape similar to all the other compounds studied. These spectra are dominated by very intense "white line" (WL) feature due to transitions into nearly empty final d states. The WL feature at Re L$_3$-edge is asymmetric, with a tiny bulge at around 3eV below its maximum, which can be attributed to the splitting of the $t_{2g}$ and $e_g$ final states [12]. Due to low resolution of the spectra, caused mainly by the core hole lifetime broadening, we did not observe any reasonable difference in their shape, which might be introduced by lattice distortion. The only alteration resolved quantitatively is a gradual decrease of the first moment of the Re L$_3$-edge spectra by 0.4eV from ReO$_3$ to Ba$_2$FeReO$_6$ and further by 0.15eV and 0.3eV for Sr$_2$FeReO$_6$ and Ca$_2$FeReO$_6$, respectively. These results indicate that all the compounds studied reveal effective valence state lower than Re$^{6+}$, that decreases with introducing smaller A-site ions, what is in agreement with Fe valence evolution observed by Mösbauer spectroscopy [6,8,16]. Nevertheless, due to the simplification in the method of analysis and strong covalent effects in these compounds, quantitative assignment of the formal valence would be unreliable.

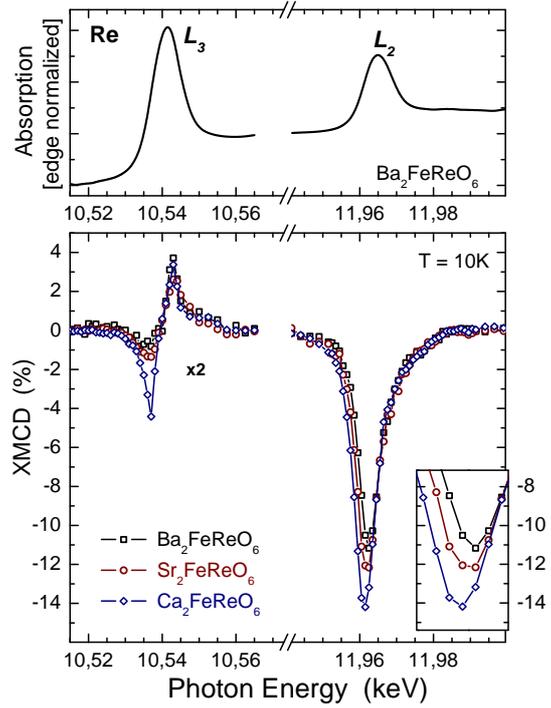

*Fig. 1.* X-ray absorption spectra of $Ba_2FeReO_6$ (upper panel) and normalized XMCD spectra (lower panel) of $Ba_2FeReO_6$, $Sr_2FeReO_6$ and $Ca_2FeReO_6$ at $L_{2,3}$-edges of rhenium measured at $T = 10$ K. Inset shows the magnified part around the minimum of the XMCD spectra at Re $L_2$-edge.

Recently, Horng-Tay Jeng and G.Y. Guo, in their theoretical study of magnetic moments in $Sr_2FeReO_6$ [14], calculated the magnetic-orbital-decomposed occupation number of Re orbitals, which indicated strong tendency to occupy the $m=1$ ($t_{2g}$) magnetic orbitals by Re electrons. Using calculated values of orbital occupation one can predict the intensity ratio of the XMCD spectra using oscillator strengths given in ref. [23]. According to such an analysis the Re $L_3$-edge dichroic spectra ought to be positive and ten time less intense than negative $L_2$-edge spectra. The latter predictions are confirmed by our experimental results, presented in Fig.1. The Re $L_3$-edge dichroic spectra show differential-like shape with negative value at the edge energy and a positive peak close to the WL maximum. The overall XMCD signal at Re $L_3$-edge is positive for all the studied compounds except $Ca_2FeReO_6$. No exception is revealed at $L_2$-edge dichroic spectra, which consist of a single negative peak. The amplitude of $L_2$-edge XMCD signal is gradually increasing with increasing lattice distortion and is approximately eight times larger than the maximum at the $L_3$-edge. Good agreement between parameters of the spectra and those predicted theoretically indicates that orbital moment is not vanishing in all the compounds studied.

In order to deduce the values of orbital, spin, and total moments from transition metal $L_{2,3}$ XMCD spectra one has to apply a set of equations known as the Sum Rules [20]. According to them, the sum of the integrated intensities under the XMCD spectra at both edges is proportional to orbital moment ($m_L$), whereas the difference of $L_3$ and doubled $L_2$ integral is proportional to the spin ($m_S$) moment and magnetic-dipole term, which can be neglected in the case of polycrystalline samples. In final relations the accurate values of moments are derived from integrated XMCD spectra, normalized to the "white line" intensity and the number of 5d holes ($n_h$). In the present study the Lorentzian shaped WL intensities were obtained by subtracting an arctan step function from the absorption spectra. The value of $n_h$ can be assumed to be between 8 ($Re^{5+}$) and 9 ($Re^{6+}$) in the compounds studied. Since the spin moments obtained for $n_h = 9$ are larger than those expected for the corresponding number of Re valence electrons ($5d^1$) the maximum number of holes "allowed" by Sum Rules ($n_{max}$) is also estimated. The spin and orbital moments derived, Tab. 1, are antiparallel, as expected from the third Hund's Rule for a less than half filled d band. The resulting total moment of rhenium is antiparallel to the Fe moment and the estimated $m_{tot} \approx 0.8\mu_B$ is in good agreement with the recent NMR [17] and neutron diffraction results [9,15]. Moreover, the spin moment of order of $1.1\mu_B$ is consistent with the results of spin density calculations [18,19].

**Table 1.** Orbital ($m_L$), spin ($m_S$) and total ($m_{tot}$) magnetic moments of rhenium at $T=10K$ in AA'FeReO$_6$ series. Values are given in $\mu_B$/f.u. for the assumed number of 5d holes ($n_h$). The $n_{max}$ parameter denotes maximum "allowed" number of 5d holes per Re ion (details in the text). Note that uncertainty of $m_L/m_S$ ratio comes from statistical and systematic errors of measurements, whereas for magnetic moments also uncertainties of WL area, relative magnetization and estimation of the number of 5d holes are taken into account. The numbers in brackets indicates the uncertainty of the last digit.

| AA' | $m_L/m_S$ | $m_L$ | $m_S$ | $m_{tot}$ | $n_h$ | $n_{max}$ |
|---|---|---|---|---|---|---|
| Ba$_2$ | -0.285(5) | 0.29(2) | -1.03(6) | -0.74(7) | 8.60 | 8.93 |
| BaSr | -0.294(4) | 0.32(2) | -1.08(7) | -0.76(7) | 8.53 | 8.88 |
| Sr$_2$ | -0.308(5) | 0.34(2) | -1.11(8) | -0.77(8) | 8.45 | 8.83 |
| SrCa | -0.327(7) | 0.39(3) | -1.18(9) | -0.79(9) | 8.31 | 8.76 |
| Ca$_2$ | -0.337(9) | 0.41(3) | -1.2(1) | -0.8(1) | 8.16 | 8.72 |

The total magnetic moment of Re in the AA'FeReO$_6$ series does not alter significantly with lattice distortion, similarly to bulk magnetization [10]. However, spin and orbital contributions are both increasing with lattice distortion, strenghtening the magnetic interactions and increasing the magnetocrystalline anisotropy, irrespectively of the number of Re 5d holes assumed in Sum Rules analysis. In particular, the increase of orbital moment by 50% between Ba$_2$FeReO$_3$ and Ca$_2$FeReO$_6$ is an unambiguous indication of a strengthening of the spin-orbit coupling in distorted compounds, as suggested by De Teresa et al. [10]. Comparison of Re and Fe magnetic moments evolution to the macroscopic magnetic properties of the compounds studied revealed that the only good correlation is observed with Curie temperature. The linear dependence of $T_C$ with the spin momenta, $S(S+1)$, of iron and rhenium, inset of Fig. 2, is similar to that predicted for Heisenberg-like exchange interactions.

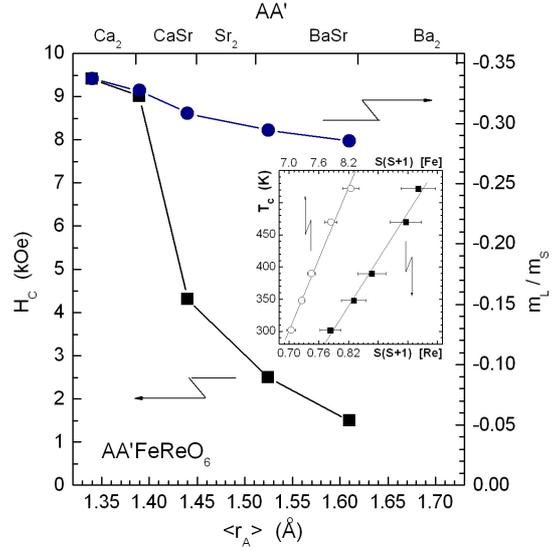

*Fig. 2. Orbital-to-spin moments ratio of rhenium (dots) and coercive field (squares) plotted versus average ionic radii of A-site alkali earths in the series of AA'FeReO$_6$ double perovskites. Inset shows the dependence of $T_C$ versus effective spin moment of Fe (open circles) and Re (filled squares) ions. The effective spin moment of Fe is derived from rhenium $n_h$ parameter with an assumption of high spin state and total number of Re and Fe d-band electrons equal to seven.*

In view of an uncertainty of absolute values of $m_L$ and $m_S$, the spin-orbit coupling is represented in our analysis by the ratio of orbital-to-spin moments. This ratio can be easily derived from the "as measured" XMCD spectra. The evolution of $m_L/m_S$ with average ionic radius of A-site alkali earths is presented in Fig. 2 together with the evolution of other unique magnetic property of Re-based double perovskites - high coercive field. Although the values of both parameters increase with lattice distortion (decreasing $<r_A>$), the evolution of the orbital contribution is less steep comparing with that of the coercive field on moving to monoclinic compounds. It indicates that nonvanishing orbital moment of rhenium is not the sole source of large intrinsic magnetocrystalline anisotropy here. A possible additional contribution can be related to an orbital moment of iron, which was predicted by spin density calculations in Sr$_2$FeReO$_6$ [14].

We addressed this problem in preliminary XMCD measurements at Fe L$_{2,3}$ edges. The results of measurements are presented in Fig. 3 together with integrals of the dichroic spectra. The presented XMCD spectra, normalized to the L$_3$ edge area, show a considerable difference of their integrals over both edges, which is a direct indication of difference in $m_L/m_S$ ratio. Although quality of the spectra is rather low, derivation of $m_L/m_S$ ratios was possible and is estimated to 0.02(3), 0.04(4) and 0.10(4) in BaSrFeReO6, Sr$_2$FeReO$_6$ and Ca$_2$FeReO$_6$, respectively. The latter results suggest the existence of a small positive orbital moment of Fe, which is consistent with more than half-filled Fe 3d band, and agrees quantitatively with the results of spin density calculations of Sr$_2$FeReO$_6$ magnetic moments [14]. A considerably larger Fe orbital moment in Ca$_2$FeReO$_6$ suggests that the spin-orbit coupling of Fe 3d electrons may be a source of the larg magnetocrystalline anisotropy in the monoclinic compounds. Nevertheless more detailed study of spin and orbital moments of iron are necessary in order to unambiguously solve this problem.

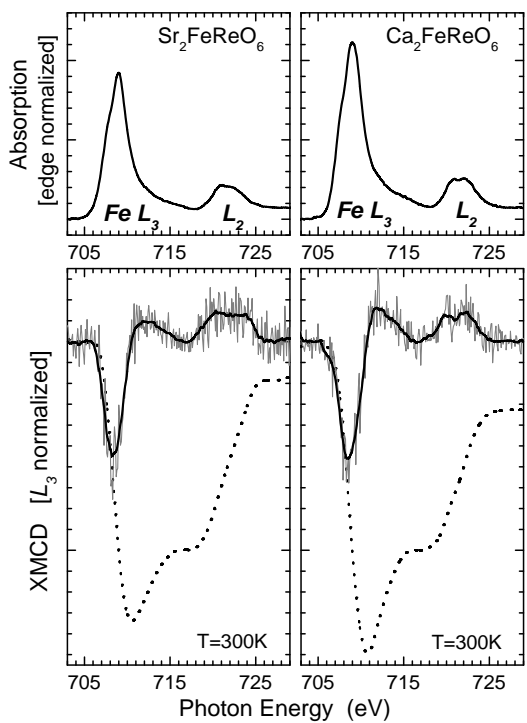

*Fig. 3. X-ray absorption (upper panels) and XMCD (lower panels) spectra of $Sr_2FeReO_6$ and $Ca_2FeReO_6$ at the $L_{2,3}$-edges of iron measured at room temperature. In lower panels the thin solid lines denote "as measured" XMCD spectra, whereas thick lines present their smoothing in the range of 1eV. Dotted lines represent integrals of dichroic spectra used for normalisation and Sum Rules analysis (see the text for details).*

In summary, detailed investigation of X-ray Magnetic Circular Dichroism spectra confirmed theoretical predictions that rhenium orbital moment is not quenched in octahedral ligand field in AA'FeReO$_6$ double perovskites. The magnetic moment of Re is of order of 0.8$\mu_B$ and is antiparallel to that of Fe, which agrees with theoretical predictions, bulk magnetization measurements and NMR spectroscopy results. The effective total angular momentum of rhenium does not change significantly within the series, whereas the magnitude of the orbital contribution increases with the lowering of local symmetry of the compounds. However, a large magnetocrystalline anisotropy in monoclinic compounds can not be explained solely by the Re orbital moment. An additional contribution from the Fe orbital moment is concluded from the Fe $L_{2,3}$ edges XMCD study.

**Acknowledgements**

We thank J. Blasco and D. Serrate (CSIC-Universidad Zaragoza) for providing the samples and the information on the magnetostriction results prior to publication as well as E. Welter (Hasylab), M.Zangrando and F.Bondino (Elettra) for their kind assistance during X-ray absorption experiments. Work supported by the State Committee for Scientific Research, Poland, Grant No. 2P03B-08223 and MAT2002-04657 funded by the Spanish Ministry of Science and Technology. Partial support by the European Commision within the frames of SCOOTMO programme, HPRN-CT-2002-00293 as well as the Research Infrastructure Action under the FP6 "Structuring the European Research Area" Programme (through the Integrated Infrastructure Initiative "Integrating Activity on Synchrotron and Free Electron Laser Science".)